\documentclass[times]{TRR}

\usepackage{moreverb,url}

\newcommand\BibTeX{{\rmfamily B\kern-.05em \textsc{i\kern-.025em b}\kern-.08em
T\kern-.1667em\lower.7ex\hbox{E}\kern-.125emX}}

\usepackage{algorithm}
\usepackage{algpseudocode}
\usepackage{amsmath}
\usepackage{booktabs}
\usepackage{multirow}
\newtheorem{definition}{Definition}
\usepackage{courier}
\usepackage{xcolor}
\definecolor{dkgreen}{rgb}{0,0.6,0}
\definecolor{gray}{rgb}{0.5,0.5,0.5}
\definecolor{mauve}{rgb}{0.58,0,0.82}
\definecolor{mygray}{gray}{0.9}
\colorlet{lightblue}{blue!70}
\definecolor{codeblue}{HTML}{0000A1}
\definecolor{lightred}{HTML}{FFEBEB}
\definecolor{tablegray}{gray}{0.9}

\usepackage[T1]{fontenc}
\usepackage{listings}
\usepackage{hyperref}
\lstdefinestyle{customasm}{
  language=[x86masm]Assembler,
  frame=tb,
  basicstyle=\bfseries\footnotesize\ttfamily,
  keepspaces=true,
  numbers=left,
  numbersep=-8pt,
  numberstyle=\footnotesize\color{gray!40!black},
  moreemph=[1]{movl,cmpl,addl},
  emphstyle = {\color{red}},
  stringstyle=\color{mauve},
  keywordstyle=\color{blue},
  commentstyle=\color{dkgreen} \textit,
}
\lstdefinestyle{customc}{
  language=C++,
  frame=tb,
  keepspaces=true,
  numbers=left,
  numbersep=-8pt,
  numberstyle=\scriptsize\color{gray!40!black},
  basicstyle=\bfseries\footnotesize\ttfamily\fontfamily{pcr},
  emphstyle={\color{red}},
  stringstyle=\color{mauve},
  keywordstyle=\color{blue},
  commentstyle=\color{dkgreen} \textit,
}

\newcommand{\ie}{\textit{i}.\textit{e}.}
\newcommand{\eg}{\textit{e}.\textit{g}.}
\DeclareMathAlphabet\mathbfcal{OMS}{cmsy}{b}{n}
\usepackage{todonotes}
\usepackage{pifont}

\begin{document}

\date{}

\title{Detecting Avalanche Effect in Adversarial Settings: Spotting the Encryption Loops in Ransomware}

\author{
Nanqing Luo\affilnum{1}, 
Xusheng Li\affilnum{2},
Haizhou Wang\affilnum{1},
Shuangyi Zhu\affilnum{3},
Yuan Ma\affilnum{4},
Peng Liu\affilnum{1}
}

\affiliation{
\affilnum{1}Pennsylvania State University\\
\affilnum{2}Vector 35 Inc.\\
\affilnum{3}Agricultural Bank of China \\ 
\affilnum{4}University of Chinese Academy of Sciences \\
}

\maketitle

\subsection*{Abstract}
Spotting encryption loops in binary-only ransomware is a critical reverse engineering task.
Since the existence of avalanche effect, an intrinsic characteristic of any secure encryption algorithms, is {\em unavoidable} during a victim data encryption attack, it is a very promising direction to spot encryption loops through avalanche effect detection.
Unfortunately, no existing work in this direction ensures that the being-checked effect is the avalanche effect {\em itself}.
Although CipherXRay \cite{li2012cipherxray} is inspired by avalanche effect, it only checks whether a ``ripple effect'' (\ie, a necessary but non-sufficient condition) of avalanche effect exists, allowing a straightforward counterattack to succeed. In this work, 
we present a new approach that checks the 
avalanche effect itself. Because the detection is conducted in 
adversarial settings (\eg, the ransomware author may obfuscate the code), a viable approach 
must tolerate inaccurate input \& output identification and must be resilient to 
adversarial evasion. These challenges are addressed by 
a novel record-and-replay detection mechanism that takes advantages of the 
statistical guarantees provided by the Shapiro–Wilk normality test.  
The experiment results show that our approach achieves achieves 0.0\% false negative rate 
and 1.1\% false positive rate. When our tool is employed to reverse engineer real-world ransomware samples, it succeeds in analyzing all the ransomware samples selected from ten representative families.

\section{Introduction} 
\label{sec:introduction}

Ransomware attack is a massive cyberattack that encrypts victims' data for ransom. 
\texttt{WannaCry} \cite{cert2017wannacry} in May 2017 is the first high-profile ransomware attack. 
\texttt{WannaCry} leveraged ``EternalBlue'', a highly effective exploit against Server Message Block (SMB), a protocol
that allows machines to share files over a Windows network. 
Within 30 days, this worldwide attack had compromised over 200,000 systems. 
In recent years, ransomware attacks have become an increasingly more serious 
societal threat due to not only the increasing amount of security loss caused by real-world 
enterprise networks, but also the increasing amount of disruptions caused to critical 
infrastructures. For example, the Colonial Pipeline ransomware attack \cite{beerman2023review} in May 2021 
has brought one of the largest and most vital oil pipelines in the United States to 
a shutdown for one week. (The pipeline comprises more than 5,500 miles of pipes.) 

\begin{figure}[htbp]
    \centering
    \centerline{\includegraphics[width=0.5\textwidth]{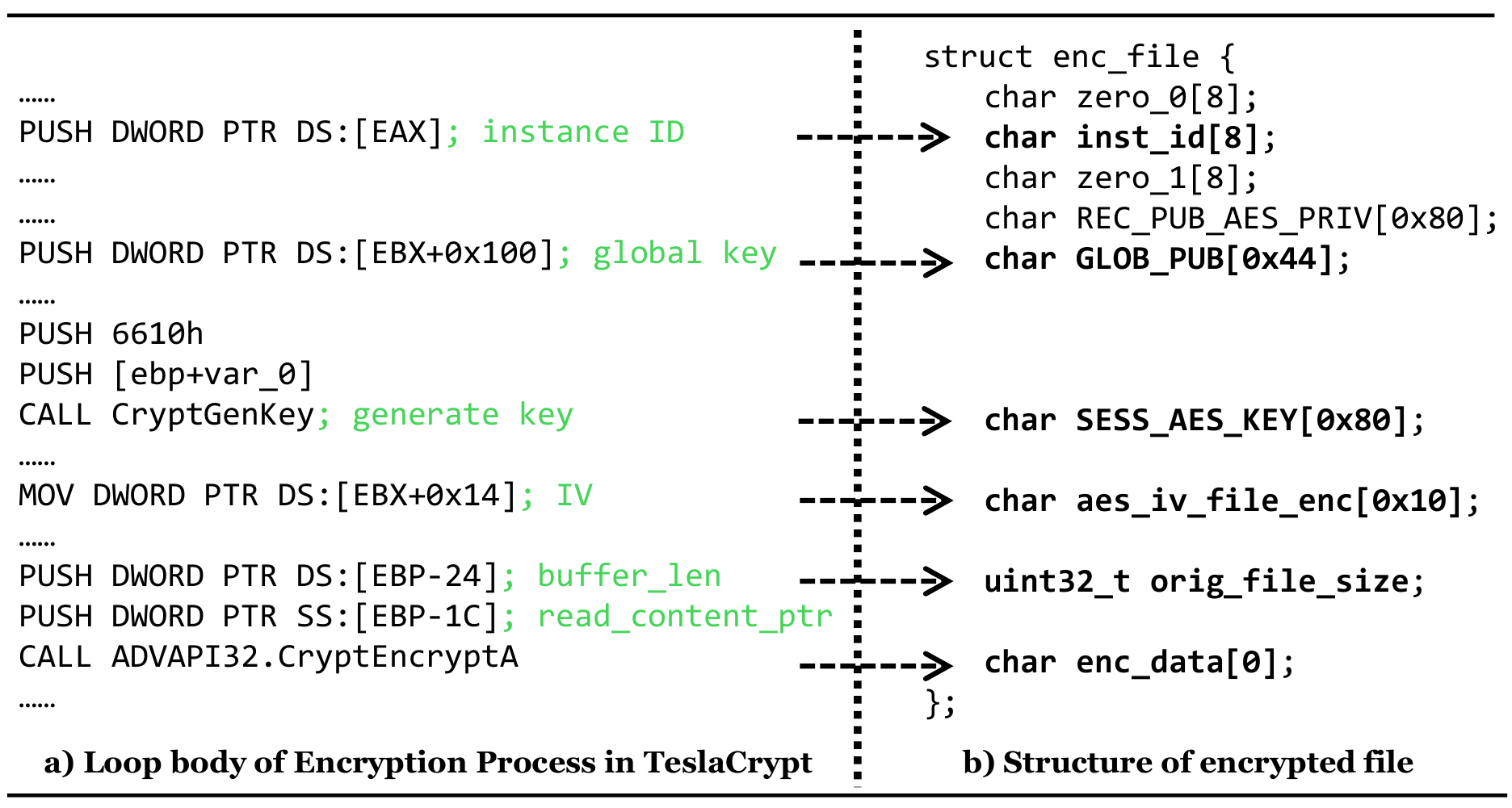}}
    \caption{Encrypted file structure analysis through loop body of encryption process in \texttt{TeslaCrypt}.}
    \label{fig:motivation}
\end{figure}

Being a critical aspect of malware analysis, reverse engineering plays an indispensable role in 
defending against ransomware attacks. 
Besides such standard malware reverse engineering tasks as 
disassembly \cite{poudyal2018framework}, type inference \cite{caballero2016type}, function recognition \cite{jia2022representation}, memory dependence analysis, 
and data/control dependence analysis \cite{banerjee1997dependence}, 
one {\em unique} task in ransomware reverse engineering 
is to identify and understand the program code 
used to {\em encrypt} victims' data.
The importance of this unique task is illustrated in \autoref{fig:motivation},
where a simplified encryption loop body of the \texttt{TeslaCrypt} ransomware \cite{skuratovich2016looking} is 
shown on the left, and the (simplified) structure of a file encrypted by \texttt{TeslaCrypt} is
on the right.
What is important here are the {\em connections} between the left side and the right side, represented 
by the directed edges.
For example, the first line of code on the left can guide 
the analyst to pinpoint the {\tt inst\_id} field in the encrypted file: this 
field tells the analyst the family/type of ransomware. 
The second line on the left can guide 
the analyst to pinpoint the \texttt{GLOB\_PUB} field: this 
field tells the analyst the public key for asymmetric encryption (\eg, RSA).
The third line on the left can guide the analyst to 
pinpoint the encrypted \texttt{AES} key in the encrypted file. 

Although this unique reverse engineering task is crucial, 
fulfilling this task in the real world 
involves arduous {\em manual labor} due to several reasons, 
including but not limited to the following ones:
(Gap A) The existing code-similarity-based methods (\eg, \cite{calvet2012aligot, xu2017cryptographic}) detect the encryption loop with high false positives. 
(Gap B) Existing pattern-based (\ie, signature/heuristic-based) methods (\eg \cite{meijer2021s, lestringant2015automated}) are not scalable when handling newly emerged encryption algorithms. 
(Gap C) Existing methods inspired by the avalanche effect \cite{li2012cipherxray} suffer from adversarial evasion which will be described in Section \textit{ \nameref{sec:motivation:cipherxray}}.

This work seeks to develop an analysis method that can 
automatically spot the encryption loops in ransomware.
Although this work is not the first avalanche-effect-inspired attempt to spot encryption loops, it is the first work that aims to 
achieve the needed resilience against {\em adversarial evasion} 
(\ie, the ransomware author can adversarially obfuscate the code 
in a variety of ways) than the existing approaches.


Besides achieving resilience against adversarial evasion, our method aims to achieve high accuracy, thus addressing the primary limitation of code-similarity-based methods, and aims to be scalable, thus addressing the primary limitation of pattern-based methods. Our method is based on two key observations: (1) The limitations identified in Gap A and Gap B can be effectively addressed through the application of avalanche effect detection, and 
(2) Gap C can be addressed by integrating statistical testing with snapshot-based replay mechanisms.

Because the avalanche effect is an {\em intrinsic} characteristic of 
any secure encryption algorithms, {\bf the existence of avalanche effect 
is unavoidable} during a victim data encryption attack. 
Informally, avalanche effect means that 
``each output bit is flipped 
with a probability of 0.50 when one bit in input is flipped''.   
In principle, if an attacker is using secure ciphers, 
the avalanche effect cannot be avoided even if 
the attacker adopted unknown ciphers or code obfuscation. 

Unfortunately, there are several daunting challenges when one intends to check 
whether the avalanche effect exists in adversarial settings. 
By ``adversarial settings'', we mean that the ransomware author is  
aware of the detection mechanism and strongly motivated to 
{\em adversarially change} the program code and/or behavior in such a way that 
confuses (\eg, triggers lots of false positives) the detection mechanism.  
First, in adversarial settings the input and output data (i.e., bits and bytes) of an encryption loop cannot be guaranteed to be identified at 100\% accuracy, and therefore the avalanche effect detection mechanism must tolerate inaccurately identified input \& output. 
Second, in order to robustly detect avalanche effect, the phenomenon of ``flipped with a probability of 0.50'' needs to be confirmed with statistical guarantees. That means we will need to reproduce this phenomenon through running the encryption loop multiple times in order to reach a statistical significant conclusion. However, no real-world malware analysis procedure takes this (statistical test) requirement into consideration.  
Third, since we are detecting avalanche effect in adversarial settings, we must achieve resilience against adversarial evasion.  
Instead of directly addressing these challenges, 
CipherXRay \cite{li2012cipherxray}, the only existing 
work in adversarial settings, takes 
the strategy of checking whether certain ``ripple effect'' (\ie, a necessary but 
non-sufficient condition) 
of the avalanche effect exists. Specifically, it checks the 
following ripple effect: if the avalanche effect exists, 
then every bit in the ciphertext should have been tainted from every bit in the
plaintext with negligible probability of exceptions. 
However, this ripple effect detection can be evaded easily:
ransomware authors can insert numerous pairs of mutually neutralizing
buffers, a specific example of data flow based counterattack, into the 
code of a ransomware sample to trigger 
CipherXRay to raise an unacceptable number of false alarms
before reporting any true positive.  
The counterattack details will be presented shortly in 
Section \textit{\nameref{sec:motivation:cipherxray}} and 
evaluated in Section \textit{\nameref{sec:eval:cipherxrayattack}}. 
(Note that this counterattack is discovered by us.) 

In this work, we ensure that what is being 
checked is the avalanche effect {\em itself}, not a ripple effect. 
To achieve this goal, a key insight is as follows: 
{\em It is possible to employ a novel record-and-replay mechanism to 
meet the specific requirements of statistical-test-based avalanche effect detection 
while tolerating inaccurate input \& output identification. }

Following this insight, our detection method has three phases.  
During Phase I, we use heuristics 
  to achieve input \& output identification. 
During Phase II, we employ the aforementioned novel mechanism to 
 check whether an output bit is crypto-related. Only 
 the crypto-related bits will enter the next phase. 
During Phase III, we conduct the Shapiro-Wilk normality test \cite{sw_test} which 
allows us to know whether an input bit is associated with avalanche effect.  
Based on the testing outcomes, we use a specific threshold 
on the minimum number of input bits that pass the test 
to determine whether the avalanche effect exists at the loop level. 

In summary, we made the following contributions: 
\begin{itemize} 
\item We present the first approach that can 
spot encryption loops in adversarial settings directly based on 
avalanche effect. 
Our approach can potentially enable the analyst 
to save a lot of time when fulfilling a critical  
ransomware reverse engineering task.  

\item We proposed a novel technique to 
ensure that what is being checked is the avalanche effect itself, 
not a ripple effect. 
In particular, we developed a record-and-replay mechanism to meet the specific 
requirements of statistical-test-based avalanche effect detection. 
The corresponding toolchain is fully implemented.  

\item The experiment results show that our tool is highly effective. 
It achieves 0.0\% FNR and 1.1\% FPR  
when the evaluation is conducted in a controlled lab environment.   
When our tool is employed to reverse engineer real-world ransomware 
samples, it succeeds in analyzing all the ransomware samples 
selected from ten representative families. On average, our tool detects 
3.2 encryption loops for each ransomware sample. 

\item We systematically analyzed the resilience of our approach 
against adversarial evasion. For example, 
when our tool is confronted with the 
aforementioned data flow-based counterattack, which can 
cause numerous failures to CipherXRay \cite{li2012cipherxray}, our tool is 
not affected at all by the attack. 
\end{itemize} 


\section{Background} \label{sec:background}

\subsection{Avalanche Effect} \label{sec:background:avalanche}
Avalanche Effect \cite{webster1985design} is one of the desirable properties of cryptographic functions with good quality. In simpler terms, it refers to the phenomenon where even slight modifications to the input of a cryptographic function result in significant changes to the output. 
Formally, if we consider a function, denoted as $f$, with an input variable of $m$ bits and an output variable of $n$ bits, the strict avalanche effect is satisfied when, for any randomly chosen inputs, flipping a single input bit will cause each of the $n$ output bits to change with a 50\% probability. 
The formal definition of the strict avalanche effect can be stated as follows:

\begin{definition} 
\label{def:avalanche}
\textbf{Strict Avalanche Effect}:
Consider two m-bits inputs $X$ and 
$X^{\prime}$, such that they differ only in bit $i$, $1 \leq i \leq m$. Let $V_{i} = Y \oplus Y^{\prime}$, where $Y = f(X)$, $Y^{\prime} = f(X^{\prime})$ and $f$ is a given  $m \times n$ function. If for all possible input pairs $X$ and $X^{\prime}$ and $1 \leq i \leq m$, the probability that each bit in $V_{i}$ equals to $1$ is 0.50, then we say function $f$ has strict avalanche effect.
\end{definition}

In the literature, it is recognized that the avalanche effect is observed not only between the plaintext and the resulting ciphertext but also between the key and the ciphertext   
in most public-key ciphers and block ciphers. However, in the case of stream ciphers, typically, the avalanche effect is only observed between the key and the expanded key stream in stream ciphers. 

\subsection{Encryption in Ransomware}
Cryptographic functions play a critical role in ransomware attacks. 
Modern ransomware typically takes a three-step encryption procedure: the first step is to generate a randomly generated RSA key by RSA algorithm with the built-in RSA key. The second step is to encrypt a randomly generated symmetric encryption key by RSA algorithm with the randomly generated RSA key. Moreover, the third step is to encrypt victim's files by a symmetric encryption algorithm with the randomly generated symmetric encryption key.

According to \cite{hull2019ransomware}, some of the most widespread ransomware in recent years, such as \texttt{CryptoLocker, CryptoWall, Locky, Cerber}, and \texttt{WannaCry}, have used block ciphers such as \texttt{AES} to encrypt the victims’ files. Alternatively, some ransomware may use stream ciphers such as \texttt{RC4} or \texttt{Salsa20} to encrypt victim files. However, stream ciphers may be more susceptible to cryptanalysis if the key stream or nonce is repeated \cite{coppersmith2002cryptanalysis}. Therefore, block ciphers have been dominating in real-world ransomware samples.
\section{Motivation, Problem Statement and Overview}

Avalanche effect is an intrinsic property 
of cryptographic functions. In the literature, it is not a new idea to use the property to assess the strength of a newly designed cipher \cite{ramanujam2011designing, kumar2012effective, vadaviya2015study, verma2020cryptography}; however, checking {\em whether the avalanche effect exists in adversarial settings} is largely uninvestigated. 
To our best knowledge, 
CipherXRay \cite{li2012cipherxray} is the only prior work that uses the property to distinguish cryptographic functions from non-crypto functions in an adversarial setting. Unfortunately, what is checked in \cite{li2012cipherxray} is {\bf not} the avalanche effect itself, but a {\em ripple effect} of the property. Since only a ripple effect is checked, 
it is not very surprising that CipherXRay can be easily evaded.
Thus, checking whether the avalanche effect exists in adversarial settings is still an open problem.  

In this section, 
we first present a counterattack against CipherXRay. Then we use the counterattack to motivate this open problem. 
Finally, we give our insights on addressing the 
new challenges and a quick overview of our approach. 

\subsection{Data flow based counterattack against CipherXRay}
\label{sec:motivation:cipherxray}

In particular, the "ripple effect" of the avalanche effect used in CipherXray is essentially a taint propagation pattern: every bit in the ciphertext should have been tainted from every bit in the plaintext with negligible probability of exceptions~\cite{li2012cipherxray}. 

Although using the taint propagation pattern is time-saving,  
what is checked is not the avalanche effect
itself.
In fact, the taint propagation pattern proposed in CipherXRay guarantees neither the presence of avalanche effect nor the presence of cryptography operations.
Therefore, encryption detection method in CipherXray is vulnerable to data flow based counterattack.
Here we propose such one data flow based Counterattack: as shown in \autoref{code:matrix}, this is a legitimate matrix multiplication loop. To CipherXRay, those are malicious crypto operations. Every byte in the result buffer \texttt{res} will be tainted by buffer \texttt{A} and buffer \texttt{B}, which 
are the two buffers inserted into the code. 
However, \autoref{code:matrix} matches the taint propagation pattern suggested by CipherXRay,
no avalanche effect exists, as we will illustrate in Section \textit{\nameref{sec:eval:cipherxrayattack}}.
Consequently, this example implies that attackers can craft 
a huge variety of mutually neutralizing buffers 
to deceive and deteriorate the effectiveness of CipherXRay with ease. 

\lstset{escapechar=@,style=customc}
\begin{lstlisting}[language=C, caption={An example of Data Flow based Counterattack.},label=code:matrix,captionpos=b]
    int main() {
        int A[3][2] = {{1, 2}, {3, 4}, {5, 6}};
        int B[2][4] = {{7, 8, 9, 10}, 
                       {11, 12, 13, 14}};
        int res[3][4];
        int i, j, k;

        for(i = 0; i < 3; i++) {
            for(j = 0; j < 4; j++) {
                res[i][j] = 0;
                for(k = 0; k < 2; k++) 
                    res[i][j] += A[i][k] * B[k][j];
            }
        }
    }
\end{lstlisting}

\textbf{Root causes of the limitation.} We observe two main causes for the limited effectiveness of CipherXRay in ransomware analysis:  \textbf{Cause 1:} Detection through taint analysis ignores values of variables, despite the fact that avalanche effect is defined upon the variations in values. \textbf{Cause 2:} Proposed ripple effect (\ie, presence of the taint pattern), albeit formally proven~\cite{li2012cipherxray}, is only necessary but not a sufficient condition of avalanche effect. Philosophically, detection through necessary conditions only is inherently vulnerable to evasion attacks. 


\begin{figure*}[htbp]
    \centering
    \centerline{\includegraphics[width=0.85\textwidth]{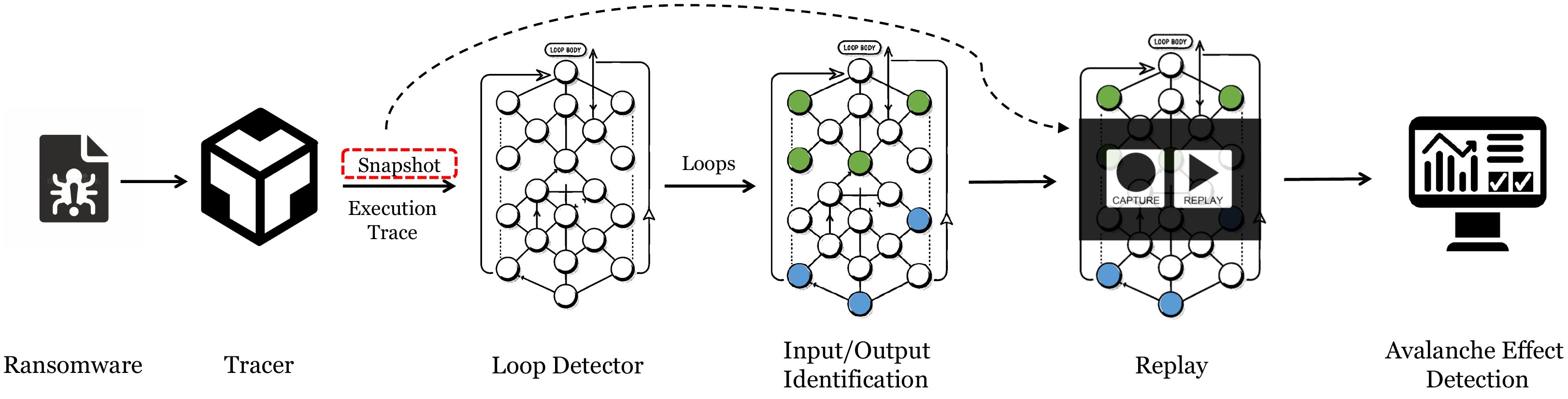}}
    \caption{Approach Overview.}
    \label{fig:overview}
\end{figure*}

\subsection{Problem Statement} \label{sec:motivation:practical}

The mutually neutralizing buffers insertion attack is shown in \autoref{code:matrix} 
against CipherXRay indicates that 
checking whether the avalanche effect exists through detection 
of a ripple effect is inherently vulnerable to evasion attacks. 

In this work, we seek to make sure that what is being checked is the avalanche effect itself. Prior to examining the specific challenges presented by adversarial settings, it is prudent to review the methodologies employed to address this problem in benign settings.

{\bf Checking whether the avalanche Effect exists in benign settings. } 
Strict avalanche effect defined in \autoref{def:avalanche} implies the probability of \textbf{each} bit in ciphertext 
being flipped is $50\%$, when one arbitrary bit in plaintext is flipped. 
Based on this definition, a practical way to check whether 
the avalanche effect exists is through measuring 
the expectation of the percentage of bits in the ciphertext. 
Formally, let's denote whether a bit at an arbitrary position $i$ of ciphertext $\mathcal{C}$ of length $n$ is flipped as $y_i \in Y$, which is an i.i.d. Bernoulli random variable with $p=0.5$. So that:
\begin{equation}
    y_i \sim Bernoulli\left(p=0.5\right)
\end{equation}
where the probability of the bit is flipped is $P\left(y_i = 1\right)$. The expectation of $y_i$ is $E\left(y_i\right) = 0.5$.
In order to check whether the avalanche effect exists, we have a new random variable describing the total number of bits flipped in the ciphertext $\mathcal{C}$ denoted as $\mathcal{A}$, so that:
\begin{equation}
    \mathcal{A} = \sum_{i=1}^n y_i
\end{equation}

Thus, the expectation of the total number of flipped bits $E\left(\mathcal{A}\right)$ is:
\begin{equation}
\begin{split}
    E\left(\mathcal{A}\right) &= E\left(\sum_{i=1}^n y_i\right) = \sum_{i=1}^n E\left(y_i\right) = \sum_{i=1}^n 0.5 = 0.5n
\end{split}
\end{equation}
Consequently, \textbf{if avalanche effect exists, the expectation of the percentage of the bits in the ciphertext that are flipped is 50\%}, because:
\begin{equation}
    \frac{E\left(\mathcal{A}\right)}{n} = \frac{0.5n}{n} = 0.5 = 50\%
\end{equation} 
Although the i.i.d. assumption may not hold for some real-world algorithms, it  aligns with the bit independence criterion of strict avalanche effect for perfectly secured cipher. Therefore, as shown in the proof above, in most real-world cases, when an analyst tests whether a cryptography algorithm bears avalanche effect, they could flip one bit in the inputs, and test whether the percentage of flipped bits in ciphertext is close to 50\%. By repeating this experiment multiple times (\eg, more than 30 input bits), a statistically significant conclusion could be reached.

{\bf Technical challenges in adversarial settings. }  
Unfortunately, measuring avalanche effect as described above 
is challenging in case of ransomware. First, during ransomware execution, the plaintext and ciphertext buffers are not guaranteed to be discovered, as the ransomware author can adopt obfuscation techniques to hide or split the buffers. 
Second, any false positive and false negative results in detecting the input and output buffers used in the encryption could be disastrous to the final measurement (\ie, the percentage of bits being flipped), which can 
lead to catastrophic failure in terms of the avalanche effect detection. 

Based on the above discussions, 
we must ensure that the solution has two properties:
\textbf{P1}: It should not assume that 
continuous input and output buffer are easily located;
\textbf{P2}: It should directly monitor variations in values 
instead of monitoring a ripple effect (\eg, via necessary condition).
Therefore, multiple executions of the cryptography operations become necessary. An intuitive solution would be similar to fuzzing: running the programs multiple times with bit-flipped inputs.
However, this intuitive solution become less practical when 
analyzing ransomware due to several obstacles:
\textbf{Obstacle 1}: The scalability is poor due to the uncertain execution time;
\textbf{Obstacle 2}: The inputs of ransomware are not controllable, and the inputs to the encryption routines are also not controllable, making the bit flipping impractical.

\subsection{Approach Overview}
To tackle the aforementioned challenges, we propose a \textit{record-replay} workflow to identify encryption operations in binary ransomware samples.
In particular, we aim to spot encryption loops. 
Spotting encryption routines at loop level is a common strategy in previous works~\cite{calvet2012aligot, xu2017cryptographic} due to several reasons. 
Loop, on one hand, is a smaller scope which can be repeatedly executed; on the other hand, loops are inevitable in any practical encryption implementation. 

As shown in \autoref{fig:overview}, our workflow includes the following important steps: Ransomware samples are first traced to get instruction level execution trace and its corresponding snapshot (\eg, \texttt{.dll}, \texttt{.data}). Then, loops are detected from the execution trace, generating execution traces for each loop (\ie, loop traces). Given loop traces, a set of inputs and outputs are pinpointed, using heuristics of memory access patterns. Loop traces are then replayed multiple times using different inputs. (Note that due to adversarial code obfuscation, we no longer know the mapping relationships between the pinpointed 
inputs/outputs and the plaintext/ciphertext.)  
Since only the {\bf loop body} is replayed, we restore the snapshot of the machine state just before the execution of the first instruction in the loop to make replay much more efficient and stable. For each round of replay, inputs are varied by flipping only one bit, and the changes in outputs are recorded. Finally, recorded change in outputs will be further analyzed using statistical testing to determine whether avalanche effect exists in each loop. 

\section{Methodology} 
\label{sec:method}

\subsection{Recording Execution Traces} \label{sec:method:trace}
The ransomware samples are executed in a customized \texttt{PyReBox} Sandbox \cite{pyrexbox}.
We have developed a tracer based on \texttt{PyReBox} which will record all the instructions executed.
For each individual instruction, The trace logs are collected in the following format:
\begin{lstlisting}
    Trace := <address>;<instr>;<raw bytes>;
             <EAX>,<EBX>,<ECX>,<EDX>,<EBP>,<ESP>;
             <read addr>,<read size>,<read val>;
             <write addr>,<write size>,<write val>;
\end{lstlisting}
where \texttt{<instr>} is the disassembled instruction (\eg, \texttt{push ebp}).  \texttt{<raw bytes>} denotes the binary data of the instruction (\eg, \texttt{0x55}).
\texttt{<EAX>} and other register tags denote the values of the registers \textbf{before} this instruction is executed.
The rest of the instruction log is memory operations, including the address, size and the values of the memory read \& write by the instruction.

Additionally, we dump all the loaded modules along with their names, size, and base addresses.
This enables us to remove those candidate loops whose instructions are all inside known system libraries that do not contain any crypto functions.
In addition, the data sections (\eg, \texttt{.data}, \texttt{.rdata}) will be loaded into the memory before the replay as part of our snapshot reconstruction. 

\subsection{Loop Detection} \label{sec:method:loop}
Some previous works detect loops by investigating repeated instruction sequences in the execution trace~\cite{calvet2012aligot, xu2017cryptographic}, which would be effective for simple loops without any branches.
However, this strategy could produce many false negatives when trying to identify loops in real-world programs, including cryptography libraries.
This is primarily due to the presence of branching statements within loops: different branches result in different instruction sequences. 

In lieu of the loop detection method used by previous works, we detect the loops by monitoring basic blocks that are previously executed in the execution trace. Since this method detects loops at basic block level, not only faster it is, but also can detect loops that have taken different branches during the execution. The detail of our loop detection method is shown in \autoref{alg:loop_detect}.

\begin{algorithm}
\caption{Loop Detection Algorithm} 
\label{alg:loop_detect}
\begin{algorithmic}[1]
\Require $B \gets \{b_1, b_2, b_2, ...\}$ \Comment{Executed Basic Blocks}
\State $S \gets \{\}$ \Comment{Stack of Frames}
\State $L \gets \{\}$ \Comment{Loop Head Results}
\ForAll {\textit{$b$ in $B$}}
    \If{$b$ \textit{ends with \texttt{ret}}}
    \State \textit{$S.pop()$}
    \ElsIf{$b$ \textit{ends with \texttt{call}}}
    \State $F \gets \{b\}$ \Comment{New frame}
    \State $S \gets S.push(F)$
    \EndIf

    \State $F \gets $ \textit{$S.top()$}
    
    \If{\textit{$b$ in $F$}} \Comment{Loop Detected}
         \State $L \leftarrow \{b, L\} $ \Comment{Save Detected Loop}
        \While{\textit{$F.top() \neq b$}}
            \State \textit{$F.pop()$}
        \EndWhile
    \Else
        \State $F.push(b)$ \Comment{Push $b$ to $F$}
    \EndIf
\EndFor
\State \textbf{return} $L$
\end{algorithmic}
\end{algorithm}

In \autoref{alg:loop_detect}, the input is a sequence of basic blocks \textit{B} executed in the trace.
The basic blocks are identified by partitioning the trace at branching instructions (\eg, \texttt{jmp, call, ret}). In order to detect loops with function calls in the loop body, the stack of frames $S$ is introduced to save stack frames $F$, which contains executed basic blocks $b$. As shown in line 4-9, if the basic block ends with \texttt{ret}, stack frame $S$ will pop the top basic block $b$. If the basic block ends with \texttt{call}, stack frame $S$ will append the basic block $b$ to contain basic blocks of the callee function.

Based on the previously mentioned basic block identification method (\ie, partitioning at branching instructions), potentially there exist two basic blocks that end with the same instruction but start with different instructions.
In the context of loop detection, these basic blocks should be considered as one basic block. Thus, the loop detector can detect a loop with different branches. Details are shown in line 10-18.
If the basic block $b$ exists in the new stack frame $F$, a loop is detected in this case.
Then, it saves the detected loop header to $L$ and pops all the basic block $b$ until it is the loop header that is previously found (line 13-15).

\subsection{Input \& Output Identification} \label{sec:method:io}
In order to check whether the avalanche effect exists in a loop, we need to examine the relation between the inputs and outputs of the loop.
Therefore, the first step is to identify the inputs and outputs of a loop.
Since our target loops are encryption loops, we assume both inputs and outputs of the loops are all memory locations.
Thus, one intuitive strategy is to first pinpoint the locations and size of the buffers, and then conduct taint analysis to find which ones are inputs and which ones are outputs, respectively. 
However, buffer detection in binary cannot be taken for granted, especially under adversarial scenarios.

Instead, 
we propose a set of heuristics based on a \textbf{three-step filter} to identify the inputs and outputs. 
The idea is that we first find a superset of 
the inputs and outputs, then gradually remove some irrelevant bytes inside them.
Specifically, we identify inputs and outputs using 3 filters: \textbf{Filter 1:} all memory read are inputs and all memory write are outputs; \textbf{Filter 2:} all constant values during the executions are removed; \textbf{Filter 3:} all obvious pointer values (\eg,~used a dereferencing memory locations) are excluded.

\lstset{escapechar=@,style=customasm}

\begin{lstlisting}[language={[x86masm]Assembler}, emph={},  caption={A section of \texttt{AES} encryption \texttt{x86} Assembly code.},label=code:asmAES,captionpos=b]
    movzx   esi, byte [kptr+4*KS_LEN] ;input pointer
    lea     tptr, [rel enc_tab]       ;load S-Box
    mov     eax, [esi+0*4]            ;input
    mov     ebx, [esi+1*4]            ;input
    mov     ecx, [esi+2*4]            ;input
    mov     edx, [esi+3*4]            ;input
    mov     ebp,[esp+ctx+stk_spc]     ;key pointer
    movzx   edi,byte [ebp+4*KS_LEN]   ;load key
\end{lstlisting}

\paragraph{Filter 1.} We first scan the memory operations of all the executed instructions in a loop trace, marking all the bytes that have been read from as input bytes, and all the bytes that have been written to as output bytes.
\autoref{code:asmAES} shows an example of detecting input in a \texttt{AES} encryption loop.
Note that a byte could be both input and output.
This is particularly common when the crypto function performs in-place encryption, \ie, the crypto function reads the plaintext, performs cryptographic transformations on it, and writes the ciphertext to the same buffer.
The marked input and output bytes will form candidate sets of input and output, respectively.

In such way, the identified input/output bytes are very inclusive.
For example, in \autoref{code:asmAES}, regardless of local or global, all read variables, including values from \texttt{S-Box}, keys, etc., are considered as input candidates.
The advantage is that it is not likely we are going to miss any data that is related to the avalanche effect, including plaintext, keys, ciphertext, etc. 
The disadvantage is that many bytes that are irrelevant to the avalanche effect may be included.
For example, the loop counter \texttt{i} is marked as both output and input because it will be compared with a target value and updated during each iteration.

\paragraph{Filter 2.} In the candidate set of input, there could exist bytes holding constant values.
In principle, those bytes with constant values should not be altered during the replay, as altering them could cause unexpected consequences.
To remove the constant values, we correlate the input/output information across all the iterations of a loop. If an input memory byte always holds the same value, we consider it to be a constant and remove it from the candidate set.
For example, in previous filter 1, the \texttt{S-box} entries in \autoref{code:asmAES} will be identified as an input candidate.
Since the \texttt{S-box} entries will never be changed during the execution, they will be filtered out here.
This consequence is indeed desirable, as detecting avalanche effect by modifying \texttt{S-box} entries is inappropriate.
Certainly, some of the input bytes, such as input bytes in the plaintext buffer and key buffer, could appear to be constant from the perspectives of a loop, but as long as the data in those buffers are transferred to other memory locations during the execution, there will exist non-constant input bytes holding the data being detected.

\paragraph{Filter 3.}
Instructions access memory to read not only data to be processed, but also pointers. Keep in mind that our goal is to detect avalanche effect, and therefore altering pointers has no merit at all.
In fact, in case pointers are identified as inputs, changing them in our replays may produce negative effects.
For example, altering a function pointer will directly lead to replay failure.
Therefore, we need to correctly

In particular, since our experiments are all conducted on X86 architecture, we adopted a brute-force heuristic: if a memory read that is 4-byte long and is never changed throughout different iterations of the loop, then this memory location will be a pointer suspect, and therefore will be excluded from the input candidate set.
Certainly, this will produce false positives and false negatives, but as long as there are inputs related to avalanche effect left in the input candidate set, our method is not affected.

As with any heuristic-based method, there would be corner cases that our heuristics cannot handle, but fortunately, this exclusion strategy will at least include enough of the true positive input and output bits for our avalanche effect detection to be functional.
This way, as long as our avalanche effect detection method can tolerate some false positive input and output bits, the avalanche effect in the loop will not be omitted.

The three filters built using heuristics mainly focus on encryption loops, based on two assumptions: 1) input and output locations are designated 
before executing the loop body, and 2) output locations are 
independent of the input values.
Therefore, our input and output identification is nothing close to perfection, but it will work reasonably well for the detection of encryption loops in ransomware, as will be shown in our experiments.

\subsection{Snapshot-based Replay}
As shown in \autoref{fig:overview}, 
upon identifying the input and output bytes, 
our workflow will enter the Replay stage.
The replay of the loop will start at the snapshot just before the start of the loop, which is reconstructed using the traces.
Since the replay is at loop level, we adopted \texttt{Unicorn} CPU emulator~\cite{unicorn}.

\paragraph{Snapshot Construction.} 
Without a correct snapshot of the program state before the execution of the loop body never can avalanche effect be detected, because an incorrect snapshot may lead to different input values (\ie,~\texttt{S-box}), causing the vanishing of avalanche effect.
Reconstruction of snapshot involves both the CPU context and memory.
Restoring the CPU context is not complicated, as the general register values are saved before the execution of each instruction.
However, restoring memory is harder, as memory locations accessed may vary from the original execution when inputs are mutated.
To restore the memory, we mimic the executable loading process, so that all the sections in the loaded images (\ie,~PE files for Windows Ransomware) will be loaded to the proper memory location based on the base addresses recorded during the execution. Then, we update the memories based on the execution trace until the first instruction of the target loop.

\paragraph{Trace Replay}
For the purpose of detecting avalanche effect, we replay the loop in the CPU emulator with different inputs.
The replay process is rather straightforward as we have all the instructions recorded except for the instructions executed in kernel mode. Thus, without any special handling will not system calls be functional.

We have only handled some common system calls such as memory allocation, so that it will no longer allocate a new buffer, but always return the same memory address as it did during the tracing.
Nevertheless, this is not a great concern as our replay is only at loop level, where not many system call will be used.
During our experiment, we rarely have replay failure due to the improper handling of system calls.

\subsection{Avalanche Effect Detection} \label{sec:method:avalanche}
As discussed in Section \textit{\nameref{sec:motivation:practical}}, given a loop with input and output identified, the most intuitive method to check avalanche effect would be flipping each input bit and then checking whether about 50\% of the output bits are flipped. 
Could this phenomenon be observed for different input bits, the avalanche effect is found.
However, this intuitive method faces two challenges: 
(\textbf{Challenge 1}) Our input and output identification heuristics make no guarantee that the identified bits are crypto-related.
Accordingly, the intuitive method may suffer from false negative: if an input bit is not crypto-related, flipping it will not render avalanche effect in the output bit; if an output bit is not crypto-related, then the percentage of bits flipped when input changes will stray away from 50\%.
(\textbf{Challenge 2}) Even if input and output bits are all crypto-related, a quantitative method  
is still missing to evaluate the fact that ``about 50\% of the bits are flipped''.

Thus, a desirable avalanche effect detection method for loops should work well even if 
not all the identified input and output bits are encryption-related.  
Furthermore, due to the inherent randomness of output bit flipping in the presence of avalanche effect, the 
detection method should be based on statistical tests. We have 
formalized our avalanche effect detection algorithm in \autoref{algo:test}, whose details 
are elaborated in the following subsections.

\begin{algorithm}[h]
  \caption{Avalanche Effect Detection}
  \begin{algorithmic}[1]
  \Require
      Input bits $I$, Output bits $O$, Threshold $\theta$
    \For{n $\in [1,30]$}
      \State $X_n \leftarrow RandomInit(I)$;   \Comment{Random values to input bits}
      \State $StopState \leftarrow Replay(X_n)$ 
      \State $Y_n \leftarrow ReadOutput(StopState, O)$ \Comment{Output values}
    \For{each $i \in I$}
      \State $(X_n)^i \leftarrow Flip(X_n, i)$;  \Comment{Input with bit $i$ flipped}
      \State $StopState \leftarrow Replay((X_n)^i)$;
      \State $(Y_n)^i \leftarrow ReadOutput(StopState, O)$
    \EndFor
    \EndFor
    
    \State
    \For{each $i\in I$}
      \State $(O)^i \leftarrow AvaOutputBits(O, \{(Y_1)^i, (Y_2)^i, ..., (Y_{30})^i\})$
      \State $(c_n)^i \leftarrow CountFlippedBits(Y_n, (Y_n)^i, (O)^i)$;
    \EndFor
    
    \State
    \State $Count \leftarrow 0$ \Comment{Avalanche Input Bits Counter}
    \ForAll{$i\in I$}
      \State $isNormal \leftarrow NormalTest(\{(c_1)^i, (c_2)^i, ..., (c_{30})^i\})$
      \If{$isNormal$}
        \State $Count \leftarrow Count + 1$
      \EndIf
    \EndFor

    \State
    \If{$Count >= \theta$}
        \State \Return Avalanche Effect Detected
    \EndIf
    \State \Return Avalanche Effect Not Detected
  \end{algorithmic}
  \label{algo:test}
\end{algorithm}

\subsubsection{Eliminating non-crypto output bits}
As stated in \textbf{Challenge 1}, the input bits and output bits identified in a cryptography loop are not necessarily all crypto-related.
Thus, before trying to detect avalanche effect, it is necessary to exclude output bits that are not crypto-related as many as possible.
According to the \textit{Strict Avalanche Effect} defined in \autoref{def:avalanche}, only in about half of the replay trials should each output bit be flipped.
Based on this criterion, after completing all the replay trials, we exclude all the output bits that 
are flipped either too often or too seldom, so that only the rest of the output bits considered in further analysis.
The exclusion strategy is rather conservative: those output bits whose 
number of times flipped falls outside the 
interquartile range of number of trails (\ie,~between 7 to 23 times for 30 trials) are excluded.
This exclusion step is reflected in \autoref{algo:test} at line 13.

\subsubsection{Testing input bits individually}
To address the other aspect of \textbf{Challenge 1}, specifically to eliminate the effect of \textit{input} bits that are not crypto-related, we focus on each input bit separately. 
That being said, each input bit is flipped to test whether it is an \textbf{Avalanche Input Bit}, by 
replaying the loop and monitoring the output bits.
However, the problem with this strategy is that a single input bit by itself can only be flipped 
and tested once, as a single bit can only have two possible values (\ie,~0 and 1).
Unfortunately, testing the input bit only once is never robust, as with a small chance 
that approximately 50\% of the output bits are flipped in that one replay run.
Besides, very few output bits that are not crypto-related could remain after the previous effort, which will further undermining the robustness of the test.

Therefore, each input bit has to be flipped multiple times, and accordingly we have developed a replay strategy to achieve this.
Although a particular input bit 
can only be flipped only once when the input value is fixed, 
we can use different input values to test the impact
of the single bit multiple times against different loop executions. 
Specifically, we replay each loop instance 30 times with different input values that are randomly initialized during each replay trial.
The reason for using 30 as the number of replay trials is related to the trade-off between 
analysis time and the prerequisite of leveraging 
the Central Limit Theorem (CLT), whose detail is described later.
After each trial of replay, we {\bf count the number of output bits flipped}.
This replay strategy is shown in \autoref{algo:test} from line 1 to line 10.

\subsubsection{Finding avalanche input bits via statistical test}
The situation described in \textbf{Challenge 2} is a perfect use case for a statistical test: we want to test whether ``about 50\% of the output bits are flipped'', given 30 sampled data points (\ie, number of output bits flipped) from flipping an input bit $i$ and replay for 30 times. 
Recall that, as discussed in Section \textit{\nameref{sec:motivation:practical}}, each output bit flipped (in the presence of avalanche effect) follows the Bernoulli distribution; therefore, the number of bits flipped in the output will follow the Binomial distribution, since Bernoulli distribution is a special case of Binomial distribution when the number of trials is 1.
Thus, testing whether ``about 50\% of the output bits are flipped'' in 30 trials is equivalent to testing the sampling distribution of the mean whose value is half of the total number of output bits, if and only if we can assume the 30 sampled data points are i.i.d Binomial random variables (and thus they have same mean).
According to the CLT, when the sample size is large enough, regardless of the underlying distribution of 
the i.i.d. random variables, the sampling distribution of mean shall follow \textbf{Normal distribution}.
In other words, if the sampling distribution of mean does not follow  
Normal distribution, 
they are not i.i.d. random variables.   

Therefore, we test the normality of the counted number of output bits flipped for each input bit $i$ flipped, as shown in \autoref{algo:test} at line 18 to line 23.
We adopt Shapiro–Wilk normality test \cite{sw_test}, whose null hypothesis is data is normally distributed. 
We have set a {\tt p-value} threshold at 0.05, so that if the test yields a {\tt p-value} less than 0.05, the null hypothesis will be rejected and the data is not normally distributed. For input bit $i$, the rejection of the null hypothesis in the normality test will directly lead to a failure of the avalanche test, so that input bit $i$ is not an \textbf{Avalanche Input Bit}.

\subsubsection{Loop Avalanche Effect}
Through the proposed statistical test, we can detect 
the Avalanche Input Bits. 
Due to the trade-off between analysis time and prerequisite of CLT, we can only 
conduct a limited number of replay trials (30 in our design).
Therefore, there is still a tiny chance that a non-Avalanche Input Bit pass the test.
To address this issue, we set up a conservative threshold $\theta = 8$ for the number of 
detected Avalanche Input Bits. Accordingly,   
we consider a loop to bear avalanche effect if the 
number of Avalanche Input Bits found is greater or equal to 8.
This procedure is reflected in \autoref{algo:test} from line 25 to line 28.



\section{Adversarial Evasion Of Proposed Method} \label{sec:evasion-cipherxray}

As discussed in Section \textit{\nameref{sec:method}}, the most important principle in designing our method is \textbf{measuring the real avalanche effect} instead of estimating it (\ie, CipherXRay does not really measure it but estimate it).
This principle should be followed primarily due to the following reason: {\em Measuring the real avalanche effect makes adversarial evasion least likely to succeed.} Note that 
since we are detecting avalanche effect in adversarial settings, any viable method must achieve resilience. 

Regarding why measuring the real avalanche effect makes adversarial evasion least likely to succeed, our first observation is as follows: Although avalanche effect is in theory an unmanipulatable property of encryption loops, a ``ripple effect'' of avalanche effect is usually 
manipulatable. For example, the 
``mutually neutralizing buffers'' attack against CipherXRay 
allows ransomware authors to manipulate a major  
ripple effect. 

Regarding why measuring the real avalanche effect makes adversarial evasion difficult, our second observation is as follows: As long as real 
avalanche effect is correctly measured, detection of 
encryption loops is guaranteed. Therefore, the aforementioned principle forces the attacker to find ways to 
``poison'' the measuring process instead of what is 
measured. 

In the remaining of this section, we will discuss the possible ways to ``poison'' the measuring process and explain why such ways are {\bf not} likely to become a viable counterattack. 
To poison the avalanche effect measuring process, 
the primary goal is to produce unwanted positives and/or unwanted negatives. 
To generate unwanted positives or negatives, it is essentially generating the avalanche effect or hiding the avalanche effect, respectively. We will discuss both cases separately.

To produce unwanted positives, essentially the attacker needs to produce avalanche effects
using operations such as hashing, chaos (dynamic) system simulation, and deep learning inference. 
However, one may notice that all these are non-trivial operations, which require significant amount of computation power, making the overhead unacceptable for ransomware. 
(Note that real world ransomware intends to encrypt user data files as quick as possible.) 
It should be noticed that the overhead caused by producing taint patterns against CipherXRay is absolutely not comparable to the overhead caused by avalanche-effect-producing routines.
Therefore, despite the potential of injecting such avalanche effect, it is not practical in real-world ransomware.

To produce unwanted negatives, the potential strategies fall into two routes: hiding the loops; hiding the avalanche effect. 
Regarding the existence of loop, because there is no way for a ransomware author to predict how many files needed be encrypted, ransomware must encapsulate the encryption operations in a loop. 
Regarding the existence of avalanche effect, it is always true that the ransomware author can use ciphers that are less secure (\eg, 7ev3n), but using such ciphers also lower the possibility of ransom getting paid. 
Since almost all the real-world ransomware samples use 
secure ciphers, handling less secure ciphers is out of the scope of this work. 

Lastly, another factor causing unwanted positives/negatives could be implementation issues, including loops being not properly detected, input and output not being properly identified, etc. 
As we will show in Section \textit{\nameref{sec:evaluation}}, our prototype is not perfectly implemented and therefore some unwanted results are produced. However, it should be noticed that such 
unwanted results are not due to the design of our detection mechanism. A novel aspect of our design is that it finds avalanche input bits through statistical test: the corresponding statistical guarantee indicates that it is 
unlikely for input bits to be improperly identified. 


\section{Evaluation} 
\label{sec:evaluation}

To comprehensively evaluate our proposed avalanche effected detection method and its practicality for real-world ransomware, we focus on the following aspects: 1) effectiveness of the avalanche effect detection, 2) fidelity of the snapshot construction, 3) encryption loop detection in real-world ransomware, 4) resilience to proposed data flow based counterattack against CipherXray, and 5) runtime overhead.



\subsection{Effectiveness of the Avalanche Effect Detection Approach} \label{sec:evaluation:avalanche}

To evaluate the soundness and completeness of our approach for checking whether the Avalanche effect exists, we first 
created a set of benign programs, including encryption, data compression, cryptographic hash function and CRC (Cyclic Redundancy Check) calculation programs. 
The set is formed so that both positive and negative cases are included: the encryption programs and cryptographic hash functions should have avalanche effect, whereas the compression and CRC calculation programs shall have no avalanche effect. 
The created benign programs include ones that use open source libraries (\eg,~\texttt{OpenSSL}, \texttt{CryptoPP}, \texttt{Lzbench}), as well as homemade ones, without using any libraries. 

Since we have the source code for all the programs and libraries,
the statistics of our testing cases can be established using the ground truth: In total, 
there are 180 compression loops, 56 CRC loops, 825 encryption loops, and 
626 cryptographic hashing loops. 
Note that the numbers are obtained by counting the {\em loop instances} found in the execution traces. 

We used our tool to handle all the testing cases. 
Among the positive samples (\ie, loops with avalanche effect), our tool has successfully 
detected all of them,
rendering \textbf{0.0\%} false negative rate. 
\autoref{tab:ava_bit_count} shows the detailed observations 
of \autoref{algo:test} for 6 selected positive loops.
Column \# {\bf{\emph{of Input bits}}} shows the number of input bits flipped; column \# {\bf{\emph{of Output bits}}} shows the number of output bits used to count the number of changed bits (line 8 in \autoref{algo:test}); and column \# {\bf{\emph{of Input bits triggers A.E.}}} shows the number of bits passing the test in \autoref{algo:test}. 
The selected loops are of different sizes, and therefore the size of input/output differs.
For example, the first loop is an instance which encrypts the data in one block belonging to an \texttt{aes-cfb-128} program. It therefore has 128 input bits and 128 output bits, respectively. 

\begin{table}[ht]
    \footnotesize
    \centering
        \caption{Avalache Effect check per bit in the benign setting.}
    \begin{tabular}{lccc}
        \toprule
        \multirow{2}{*}{\bf{\emph{Loop}}}       & \# \bf{\emph{of Input}} & \# \bf{\emph{of Output}} & \# \bf{\emph{of Input bits}} \\
                                    & \bf{\emph{bits}}        & \bf{\emph{bits}}         &  \bf{\emph{triggers A.E.}} \\
        \midrule
        \texttt{aes\_cfb-1} & 128              & 128                & 128                            \\
        \texttt{aes\_cfb-2} & 6144             & 3768              & 6038                           \\
        \texttt{md5-1}      & 7360             & 512               & 7360                           \\
        \texttt{sha2\_512-1}   & 408              & 512               & 312                            \\
        \texttt{sha3\_512-1}   & 9408             & 11712             & 9271                           \\
        \texttt{des\_cbc-1}  & 20880            & 8128              & 20778                         \\
        \bottomrule
    \end{tabular}

    \label{tab:ava_bit_count}
\end{table}

In contrast, regarding the negative samples, our tool did not perform perfectly: Out of 236 negative samples, 3 of them are reported to have avalanche effect, rendering a \textbf{1.1\%} false positive rate.
After examining the false positives, we found that all of them are 
complicated loops, involving numerous branches (\eg, \texttt{if-else}).
In every false positive case, the identified inputs can 
significantly affect the control flow of the loop and thus the code executed.
Consequently, when these input values are modified, the output location will change.
Recall our input/output detection is based on memory operations, and 
therefore by ``output location'', we are referring to the memory locations detected to be the output. 
\subsection{Fidelity Study of Snapshot Construction}

Based on our experiment, during the replay of the loop, the avalanche effect can no longer be detected even if 5\% of the memory in the snapshot are randomized.
Therefore, one premise of our record-replay workflow is that the replays would achieve high-fidelity. 

Our idea to verify the snapshot reconstruction is as follows: 
Starting from the {\em same} reconstructed snapshot, we execute the the same loop body on not only 
a real processor but also our replay emulator.
The execution will stop at the first instruction outside the loop.
We conclude our snapshot reconstruction is high-fidelity only if the non-reserved registers (\ie, general registers and \texttt{EFLAGS} register) from both real processor and emulator are identical.

Following this idea, we created three benign programs that contain cryptographic loops (\ie,~loops with Avalanche Effect), including loops in \texttt{md5}, \texttt{des\_cbc}, and \texttt{aes\_cfb}.
For each program, we inserted 4 "breakpoints" into selected cryptographic loops and in total we have 599 breakpoints.
Using different inputs (by manipulating memory), we observe that in all 599 breakpoints, the values of all non-reserved registers in both real processor and emulator are consistent.

\subsection{Encryption Loop Detection in Ransomware}
\label{sec:eval:encdetect}

To collect ransomware traces, we build a tracer on top of the \texttt{PyRebox} sandbox. 

First, we executed the ransomware in the sandbox with {\em decoy} files deployed.
Before the tracing starts, we monitor all the ransomware processes and hook 
the API call related to file read/write (\eg,~\texttt{ZwWriteFile}).
Only after the disk read/write of decoy files is observed will we start the tracing on the corresponding process. 
This strategy minimizes the amount of tracing time.
Then, the tracer will start recording the trace as described in Section \textit{\nameref{sec:method:trace}}.
Since the encryption may be executed in a multi-thread manner, the tracer will record execution traces for all threads, generating a separate trace file for each thread.
Finally, the tracer stops when the disk write of the 
encrypted file is observed, ensuring that the recorded trace contains encryption loops.
Before the tracer stops, all loaded libraries (\eg, \texttt{.dll}) are dumped, and 
their base addresses are saved for further snapshot construction.

To save the analysis time, some loops can be skipped. 
There are several criteria to filter out a loop: 1) It is in a known system library (\eg, \texttt{user32.dll}); 2) it contains inner encryption loops identified already.

As shown in \autoref{tab:ransom}, we select 10 ransomware samples out of 10 distinct 
families of ransomware. Note that here we exclude those potential crypto-hashing loops. To filter out these hash loops, we count the average number of bitwise AND and OR operations per loop iteration, discarding loops with more than 5 such operations as they are likely to be hash loops rather than encryption loops.
For all the tested ransomware samples, all the crypto loops identified by 
our tool are correct, and zero false positives and false negatives are observed. 
On average, 3.2 encryption loops were detected for each ransomware sample. 
We manually verified every loop that our toolchain 
reported as positive.
To verify an encryption loop, we first check whether the loop is from dynamically loaded libraries.
If it is from encryption functions from libraries that contain known encryption functions (\eg, \texttt{ADVAPI32.dll}, \texttt{Cryptbase.dll}), we confirm that it is an encryption loop.
In case the loop is not from libraries, we then perform manual checks, instruction by instruction. We focus on the opcodes and memory operations.
It can be seen that real-world ransomware is more complex and challenging as 
reflected by the number of candidate loops (see Section \textit{\nameref{sec:method:trace}}, \textit{\nameref{sec:method:loop}}. 
On average, the total number of candidate loops for each ransomware sample is 35.7. 
Nevertheless, the experiments indicate that our toolchain is able to handle 
the workload. Among all the samples, we have two interesting observations: 
\textbf{Observation 1}: Most ransomware samples employ encryption 
algorithms implemented in a library (\eg, \texttt{Advapi32.dll, Crypt32.dll}).
\textbf{Observation 2}: We have identified the presence of an encryption loop within the \texttt{Avoslocker} executable (\ie, not in any library). This observation leads us to believe that the loop is associated with string encryption, as it exhibits certain 
string manipulation operations (\eg, \texttt{bswap}). 

\begin{table}[htbp]
\centering
\footnotesize
\caption{Encryption detection result in Ransomware.}
\begin{tabular}{cccccc}
\toprule
\bf{\emph{Name}} & \bf{\emph{Family}}          & \begin{tabular}[c]{@{}c@{}}\bf{\emph{N\textsuperscript{\underline{o}} of}}\\ \bf{\emph{Loop}} \end{tabular} &
  \begin{tabular}[c]{@{}c@{}}\bf{\emph{Enc.}} \\ \bf{\emph{Loop}}\end{tabular} &
  \begin{tabular}[c]{@{}c@{}}\bf{\emph{Filtered}}\\ \bf{\emph{Loop}}\end{tabular} & \bf{\emph{FP}} \\ \midrule
84c8   & \texttt{WannaCry}   & 6   & 3    & 3         & 0  \\
01c6   & \texttt{Babuk}      & 1   & 1    & 0         & 0  \\
0b1f   & \texttt{AvosLocker} & 164 & 6    & 66        & 0  \\
c249   & \texttt{HelloKitty} & 60  & 6    & 15        & 0  \\
008e   & \texttt{Makop}      & 52  & 5    & 10        & 0  \\
1a05   & \texttt{Dharma}     & 8   & 2    & 1         & 0  \\
1f1a   & \texttt{Rapid}     & 1   & 1    & 0         & 0  \\
70fa   & \texttt{GlobeImposter}     & 37   & 3    & 0         & 0  \\
8be4   & \texttt{MoneyMessage}     & 11   & 2    & 0         & 0  \\
73ca   & \texttt{ItsSoEasy}     & 17   & 3    & 4         & 0  \\
\bottomrule
\end{tabular}
\label{tab:ransom}
\end{table}

\subsection{Resilience to Data Flow Based Counterattack}
\label{sec:eval:cipherxrayattack}
One key motivation of our work is to tackle the limitation of CipherXray~\cite{li2012cipherxray}, which also leverages the Avalanche effect to detect encryption functions. 
As elaborated in Section \textit{\nameref{sec:motivation:cipherxray}}, CipherXray tries to check whether the Avalanche effect exists through taint analysis across buffers, which could be evaded in adversarial settings. 
We have proposed a data flow based counterattack in \autoref{code:matrix} that should evade CipherXray.
To test this hypothesis, we first create a program sample 
by injecting several encryption loops (\ie,~loops with avalanche effect), as well as the loops in \autoref{code:matrix} (\ie,~loops without avalanche effect).
Then we investigate whether CipherXray and our tool report 
the Avalanche effect from the injected loops.

Regarding the implementation of CipherXray, since the authors released neither the source code nor the binary of their tool, we re-implement their encryption function 
detection algorithm using Intel PIN and Triton \cite{SSTIC2015-Saudel-Salwan}. 
To collect the trace of the injected loops, we first use Intel PIN to collect the execution trace, and then extract each injected loop.
Given the trace of each injected loop, we conduct the taint analysis as described in CipherXray 
using Triton.
Based on our observation, when all the bytes in 
either buffer \texttt{A} or buffer \texttt{B} (see \autoref{code:matrix}) 
are tainted, all 48 bytes in \texttt{res} are also tainted. 
This observation is aligned with our hypothesis, and therefore, 
CipherXray shall report this loop as an encryption loop, which is a false positive.

In contrast, our tool reports no Avalanche effect from the loops in \autoref{code:matrix}. Since our method takes values into account, we repeat the experiments with different values in buffers \texttt{A} and \texttt{B}. 
Still, no encryption loop is reported, even though all 48 bytes in \texttt{res} are tainted, 

Consequently, when adopting CipherXray in an adversarial setting, if the 
attacker is aware of the detection scheme, he/she can inject hundreds of 
such loops (into the code of a ransomware sample) 
to generate substantial number of false positives, causing the 
detection procedure impractical. 



\subsection{Runtime Overhead}
The loop detection time is contingent upon the size of the execution trace.
Since our loop detector is operated at basic block level, the size of execution trace is in terms of the number of basic blocks.
As shown in \autoref{tab:overhead}, the processing time for individual loops (\ie, column \textbf{\emph{Loop Detection}}) is typically quite short, requiring no more than one second per loop, indicating a high level of detection efficiency. 

Although we replay all the detected loops in one ransomware sample multiple times, the runtime overhead is still acceptable thanks to our efficient snapshot construction. As shown in \autoref{tab:overhead}, the cumulative time consumption of detection, encompassing both loop detection and the replay phase, remains within acceptable bounds for a single ransomware sample. It becomes evident that as the trace file size increases, the time required for loop detection and replay also significantly grows.
Given our deliberate commitment to ensuring that the encryption process is comprehensively captured within the trace, the trace file has grown substantially in size.
Consequently, optimizing the tracer holds the potential to enhance the performance of both the loop detector and the replay process. However, it should be noted that the pursuit of such optimization falls outside the purview of this research.

On the other hand, owing to the streamlined and efficient snapshot construction process, the average duration for replaying (\ie, column \textbf{\emph{Avg. Replay}}) a single loop typically ranges from 1 to 5 minutes, which is considerably less than the time required for manual inspection of reverse engineers. 

\begin{table}[htbp]
\centering
\caption{Runtime overhead in seconds.}
\begin{tabular}{cccc}
\toprule
\bf{\emph{Name}} & \bf{\emph{Loop Detection}} & \bf{\emph{Avg. Replay}} & \bf{\emph{Overall}}  \\ \midrule
84c8 & 133.0            & 141.30       & 556.94      \\
01c6 & 0.3            & 3730.20      & 3730.50   \\
0b1f & 22.0             & 66.65       & 11019.29 \\
c249 & 18.0             & 336.10      & 20172.10  \\
008e & 8.0              & 82.11      & 4277.95 \\ 
1a05 & 1.1              & 28.54     & 229.42 \\
1f1a & 1.8              & 32.35     & 34.15 \\
70fa & 5.9              & 57.16     & 2121.16 \\
8be4 & 6.0              & 184.40     & 2034.43 \\
73ca & 2.7              & 78.70     & 1340.60 \\
\bottomrule
\end{tabular}

\label{tab:overhead}
\end{table}

\subsection{False Positive Case Study}
The first case pertains to cryptographic hashing loops with a reduced frequency of \texttt{and}, and \texttt{or} opcodes. Note that the crypto hashing loop also exhibits the avalanche effect. As mentioned in Section \textit{\nameref{sec:eval:encdetect}}, a heuristic approach—specifically, the calculation of the mean frequency of bitwise and and or operations per loop iteration—is employed to identify and exclude cryptographic hashing loops. However, our investigation revealed that several contemporary cryptographic hash functions have been optimized for modern CPU architectures. A salient example is BLAKE3, which is predicated on the ARX (\ie, modular Add, Rotate, and XOR) design paradigm, offering superior performance on modern CPUs. While it does not completely avoid \texttt{and}, \texttt{or} operations, it uses them less frequently than some other hash functions. The core mixing function does not use \texttt{and}, or \texttt{or} at all. Nevertheless, it is crucial to emphasize that the ineffectiveness of the hash function filter does not compromise the capability of the avalanche effect detection mechanism.

The second case is that some compression loops are wrongly detected to have the avalanche effect. Our investigation reveals that this is due to the input/output misidentification. As shown in \autoref{code:compfp}, the function \texttt{compress\_block} takes \texttt{block} as actual input, the \textit{final} \texttt{state} as the actual output, and uses a temporary buffer \texttt{temp} for intermediate calculations. It performs 10 rounds of mixing operations. Note that, the actual avalanche effect should only be measured by how this 1-bit change in \texttt{block} affects the \textit{final} \texttt{state} after all rounds and the final XOR. 
However, the inclusive input/output identification also considers \textit{initial} \texttt{state} as input, \texttt{temp}, or intermediate values of \texttt{state} as outputs. Assume that we change one bit in the input \texttt{block}, and the change propagates through the rounds, affecting \texttt{temp}. If \texttt{temp} is considered an output, large changes in \texttt{temp} during rounds might be seen as a strong avalanche effect. If intermediate \texttt{state} values are considered outputs, their changes during rounds might also be mistaken for avalanche effects. This analysis, while addressing a rare occurrence, underscores the critical importance of precise input/output identification in the evaluation of cryptographic properties.

\lstset{escapechar=@,style=customc}
\begin{lstlisting}[language=C, basicstyle=\footnotesize\ttfamily, emph={},  caption={An example Compression Function.},label=code:compfp,captionpos=b]
    #define BLOCK_SIZE 16
    #define STATE_SIZE 4
    
    void compress_block(uint32_t *state, 
                        const uint8_t *block) {
        uint32_t temp[STATE_SIZE];
        
        // Copy state to temporary buffer
        for (int i = 0; i < STATE_SIZE; i++) {
            temp[i] = state[i];
        }
        
        // Compression rounds
        for (int round = 0; round < 10; round++) {
            // Mix state with input block
            for (int i = 0; i < STATE_SIZE; i++) {
                temp[i] += *((uint32_t*)(block + 4*i));
                temp[i] = (temp[i] << 7) | (temp[i] >> 25); 
                temp[i] ^= temp[(i+1) % STATE_SIZE];
            }
            
            // Additional mixing
            for (int i = 0; i < STATE_SIZE; i++) {
                temp[i] += temp[(i+1) % STATE_SIZE];
                temp[i] = (temp[i] << 13) | (temp[i] >> 19);
            }
        }
        
        // Update final state
        for (int i = 0; i < STATE_SIZE; i++) {
            state[i] ^= temp[i];
        }
    }
\end{lstlisting}

\section{Related Work}

\paragraph{Avalanche Effect}
Researchers have strived to design and analyze encryption schemes that exhibit a strong avalanche effect, as it is a fundamental requirement for ensuring the confidentiality and integrity of data. Numerous studies \cite{ramanujam2011designing, kumar2012effective, vadaviya2015study, verma2020cryptography} have focused on evaluating and enhancing the avalanche effect in cryptographic systems to resist various attacks, such as differential and linear cryptanalysis \cite{heys2002tutorial, matsui1993linear}. Classic ciphers like the Data Encryption Standard (\texttt{DES}) and more modern algorithms like the Advanced Encryption Standard (\texttt{AES}) have been extensively analyzed and praised for their strong avalanche properties \cite{bruce1996applied, joan2002design}. It should be noticed that our work is clearly different from these works: they all evaluate and enhance the avalanche effect in {\em benign} settings; in contrast, our work has to measure avalanche effect in adversarial settings. 

\paragraph{Crypto Function Detection} Crypto functions are usually detected by both static \cite{braga2019understanding,lestringant2015automated, meijer2021s} and dynamic \cite{de2017kali,akbanov2018static,calvet2012aligot,xu2017cryptographic,li2012cipherxray,li2018k} analysis. The approaches to detect crypto functions can be split into two categories: Code-similarity-based \cite{xu2017cryptographic, calvet2012aligot} and signature/heuristic-based \cite{li2012cipherxray, de2017kali, akbanov2018static, li2018k, meijer2021s}. 
\cite{de2017kali} and \cite{akbanov2018static} trace this system execution at the production stage and detect the input-output relationships among viable pairs of crypto keys, enter buffers, and/or output buffers. \cite{calvet2012aligot} proposes Aligot to identify cryptographic functions retrieving their parameters in a way that is basically impartial. \cite{xu2017cryptographic} comes up with a novel Bit-precise symbol execution to detect crypto functions and get good performance in obfuscated programs.  
By examining program execution traces, k-hunt \cite{li2018k} discovers improperly produced keys, insecurely negotiated keys, and recoverable keys.

\section{Discussion and Limitations}

\paragraph{Discussions on anti dynamic analysis} 
To collect execution traces from ransomware binaries, the most common ways are dynamic binary instrumentation (DBI) and sandbox.
Since DBI based tool such as \texttt{Intel Pin} provide no transparency at all, we build our tracer upon the \texttt{PyRebox} sandbox.
Although it is not perfect, nearly 90\% of the ransomware samples we analyzed can be properly executed and traced by the dynamic tracer we build.
Among the rest of the 10\% ransomware samples we failed to analyze, most of them are due to inactive C2C servers.


\paragraph{Discussions on program obfuscation}
One key feature of our method is to measure \textit{real} avalanche effect so that evasion discussed in Section \textit{\nameref{sec:evasion-cipherxray}} is ineffective.
However, our method could still be vulnerable to adversarial evasions using program code obfuscations, which is a general problem for program analysis.
Fortunately, many common obfuscation schemes are {\bf ineffective} against our design, because: 1) our replay is based on traces from concrete execution, so obfuscation schemes targeting static analysis are ineffective; 2) our analysis is at loop level, and our loop detection is solid as long as the loop body had been repeatedly executed more than once.
Using the three obfuscation schemes in OLLVM (\ie, control flow flattening (FLA), instruction substitution (SUB), and bogus control flow (BCF)) as examples, we briefly explain why our method can resist them.
FLA is ineffective because it cannot hide loops.
As long as loops can be correctly identified, there is no impact on our method.
SUB is ineffective because our avalanche effect detection is based on observing the changes in the inputs and outputs, so that as long as the inputs and outputs are correctly identified, the exact instructions executed are not our concern.
BCF is ineffective simply because this is an obfuscation for static analysis so that the bogus control flow will never be taken during our concrete execution and therefore will not affect our collected traces.

\paragraph{Discussions on Input \& Output identification} While the Input \& Output identification method discussed in Section \textit{\nameref{sec:method:io}} provides heuristics to discerning inputs and outputs within binary execution traces, the lack of enough semantic information in binary and the potential presence of obfuscation techniques may challenge the accuracy of the identification process. 
In this paper, since our Avalanche input bit detection algorithm is intentionally designed 
to tolerate inaccurate Input \& Output identification, 
we adopt a heuristic-based strategy to identify input and output memory locations. 
However, as shown in the experiments, there are some corner cases and we will further improve the identification method in future work.

\section{Conclusion}

We propose a method to detect encryption loops in adversarial settings, specifically binary-only ransomware samples. The experiments show that our approach is highly effective. It achieves 0.0\% false negative rate 
and 1.1\% false positive rate 
when the evaluation is conducted in a controlled
lab environment. 
When our tool is employed to reverse engineer real-world ransomware samples, it succeeds in analyzing all the ransomware samples selected from ten representative families. On average, our tool detects 3.2
encryption loops for each ransomware sample.

\bibliographystyle{TRR}
\bibliography{sample}

\end{document}